\documentclass[prl,10pt,twocolumn]{revtex4}

\usepackage{amsmath}    
\usepackage{graphicx}   
\usepackage{verbatim}   
\usepackage{color}      
\usepackage{subfigure}  
\usepackage{hyperref}   

\begin{document}

\title{Viscous spin exchange torque on precessional magnetization\\ in $(\mathrm{LaMnO}_3)_{2n}/(\mathrm{SrMnO}_3)_{n}$ superlattices}
\author{H. B. Zhao, K. J. Smith, Y. Fan, G. L\"{u}pke}
\affiliation{Department of Applied Science, College of William and Mary, Williamsburg, Virginia 23187}
\author{A. Bhattacharya, S. D. Bader}
\affiliation{Center for Nanoscale Materials and Materials Science Division, Argonne National Laboratory, Argonne, Illinois 60349}
\author{M. Warusawithana, X. Zhai, J. N. Eckstein}
\affiliation{Department of Physics, University of Illinois at Urbana-Champaign, Illinois 60801}
\date{June 12, 2007}
\begin{abstract}
Photoinduced magnetization dynamics is investigated in chemically ordered $(\mathrm{LaMnO}_3)_{2n}/(\mathrm{SrMnO}_3)_n$ superlattices using the time-resolved magneto-optic Kerr effect. A monotonic frequency-field dependence is observed for the $n=1$ superlattice, indicating a single spin population consistent with a homogeneous hole distribution. In contrast, for $n\geq2$ superlattices, a large precession frequency is observed at low fields indicating the presence of an exchange torque in the dynamic regime. We propose a model that ascribes the emergence of exchange torque to the coupling between two spin populations -- viscous and fast spins.
\end{abstract}
\maketitle
The manganites exhibit a rich variety of electronic and magnetic phases as a function of doping, or substitution of the A-site cation [\textsl{e.g.} $\mathrm{La}^{3+}$ and $\mathrm{Sr}^{2+}$ in $(\mathrm{La}_{1-x}\mathrm{Sr}_x)\mathrm{MnO}_3]$ to allow the Mn sites to have mixed valence~\cite{R1}. An optimized carrier concentration favors the double exchange interaction which tends to globally align spins on Mn sites ferromagnetically ~\cite{R2}. However, in some doping and temperature ranges, random occupancy on the $A$-site may give rise to coexisting regions with locally ferromagnetic (FM) and antiferromagnetic (AF) correlations. In such a regime, a small magnetic field may have dramatic consequences, because it can align the randomly oriented spins and energetically promote ferromagnetism, rendering a global ferromagnetic order out of a mixed phases. 

The phase complexity in doped manganites makes the investigation of periodic layered structures especially intriguing, given the possibility of engineering charge, spin and orbital ordering in a layer-by-layer manner~\cite{R3,R4,R5}. Using ozone assisted oxide-MBE (Molecular Beam Epitaxy), and a strategy called 'Digital Synthesis', superlattices can be synthesized using integer unit-cell layers of constituent materials that are fully ordered on the $A$-site~\cite{R6}.  First, such structures may lead to novel magnetic interactions due to reconstructions involving charge, spin and orbital degrees of freedom across atomically sharp interfaces, which is absent in materials where $3^+$ and $2^+$ cations randomly occupy the $A$-site. Secondly, the coexisting phases would be regularly arranged, allowing characterization with scattering probes that exploit this periodicity~\cite{R7}, or via proximity effects. These approaches can be used to investigate the intimate correlation of the magnetic interaction with the ordered electronic structure, and how this influences the phase transitions into the various ordered states.  

Recently, ferromagnetic order has been realized in superlattices of dissimilar AF insulating manganites, $\mathrm{LaMnO}_3$ (LMO) and $\mathrm{SrMnO}_3$ (SMO), due to charge transfers promoted by the chemical potential difference across the interface~\cite{R4,R8}. The AF interaction would still be favored in other regions far from the interface, but a region with glassy/frustrated spins may exist between the FM and AF layers. The magnetic coupling between these layers can dramatically affect the magnetization switching and dynamics, which may be of interest in the context of magnetoelectronics~\cite{R9}.

Recent studies of the magnetization dynamics in ferromagnetic manganite random-alloy films have shown that the uniform spin precession mode is governed by anisotropy and demagnetizing fields~\cite{R10,R11,R12,R13}. Due to a homogeneous hole distribution in these films, spins on all Mn sites are coupled parallel by strong ferromagnetic interactions. Therefore, a single spin population is sufficient to describe both quasi-static and dynamic magnetization measurements. However, this may not be true in manganite superlattices, in which the density of holes is modulated, thus modulating the magnetic interaction and spin character in both the static and dynamic regime. 

In this Letter, we report on a study of the uniform spin precession dynamics in $(\textrm{LMO})_{2n}/(\textrm{SMO})_n$ superlattices through ultrafast pump-probe measurements of the time-resolved magneto-optical Kerr effect (TR-MOKE). We show that the one-spin model typically invoked does not adequately explain the observed frequency-field behavior for the case where $n\geq2$. A large precessional frequency observed at low fields indicates the presence of an exchange torque in the dynamic regime. In contrast, the FM superlattice with $n=1$ and the corresponding random-alloy thin film are shown to have similar precession dynamics, which can be described well by a single spin population and its anisotropy fields. We propose a model that ascribes the emergence of exchange torque in the manganite superlattice to the coupling between two spin populations - viscous and fast spins. 

TR-MOKE measurements are performed with a Ti:sapphire amplifier laser system providing 150-fs pulses at 1-kHz repetition rate. Figure 1(a) depicts the geometry of our TR-MOKE setup. We use 800-nm pump pulses with fluence of 1 mJ/cm$^2$ to induce magnetization precession in our samples, and the time evolution of the out-of-plane magnetization component $\textrm{M}_z$ is monitored with time-delayed 400-nm probe pulses with fluences on the order of 0.1 mJ/cm$^2$. A split-coil superconducting magnet is employed to study the field dependence of the magnetization precession for both in-plane and out-of-plane sample geometries. All data shown were taken at $\textrm{T}=50 K$.

Figure 1(c) shows the time evolution of the out-of-plane magnetization component, as measured by TR-MOKE for the $(\textrm{LMO})_2/(\textrm{SMO})_1$ sample with its spin structure depicted in Fig. 1 (b). All spins are coupled parallel due to strong ferromagnetic exchange interactions in this n=1 superlattice. As a result, an intense oscillation due to the uniform precession of FM spins is observed. The solid line in Fig. 1(c) indicates a fit yielding a precession frequency, $f=6.3$ GHz, and a damping rate, $\Gamma=0.0024$ ps$^{-1}$, defined by $\textrm{M}_z \approx \exp(i2\pi f t-\Gamma t)$.

The excitation and precession of the magnetization can be explained by laser-induced demagnetization and anisotropy modulation~\cite{R14,R15}. When the external field is applied normal to the sample plane, the strong pump pulse instantaneously heats up the sample, which alters the equilibrium direction of the magnetization caused by a sudden change of the demagnetization field. The magnetization subsequently rotates towards the direction of a transient field ($\mathbf{H}_{\textrm{tr}}$), and after the sample cools ($\Delta t>50$ ps), precesses around the original effective field $\mathbf{H}_\textrm{eff}$, as indicated in the inset of Fig. 1(c). 

\begin{figure}
	\centering
		\includegraphics[width=\columnwidth]{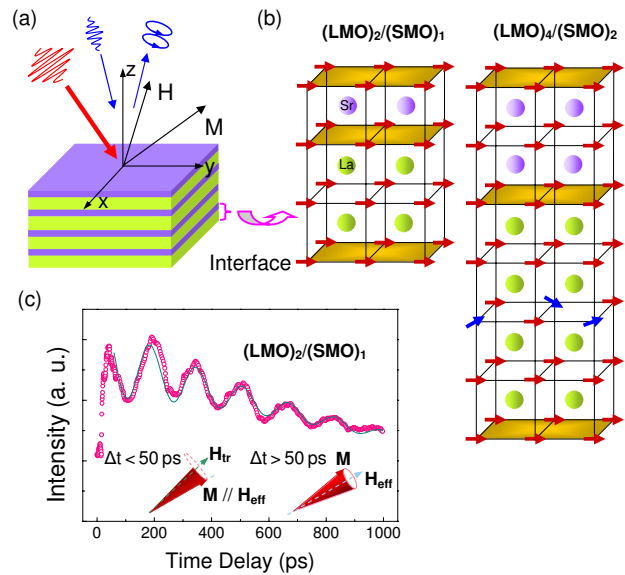}
	\label{fig:fig1}
	\caption{(a) Geometry of the TR-MOKE measurements with applied magnetic field H nearly normal to sample surface ($\approx 4^\circ$ between H and z direction); (b) Spin structure of $(\mathrm{LMO})_2/(\mathrm{SMO})_1$ and $(\mathrm{LMO})_4/(\mathrm{SMO})_2$ superlattices. Red and blue arrows represent ferromagnetic and viscous spins, respectively; (c) Time evolution of magnetization precession measured at H=1.0 T for $(\mathrm{LMO})_2/(\mathrm{SMO})_1$. The solid line is a fit to the oscillatory part, as described in the text.}
\end{figure}

Figures 2(a) and (b) show the field dependence of the uniform spin precession for the $(\textrm{LMO})_{2n}/(\textrm{SMO})_{n}$ superlattices with $n=1$ and $n=2$, respectively. The precession frequencies are plotted in Fig. 3.  The frequency of the $n=1$ superlattice clearly decreases monotonically to about zero with decreasing field. However, for the $n=2$ superlattice, there is a local minimum at $\mathrm{H}\approx0.8$ T, and an extrapolated non-zero value of $\approx 11$ GHz in the limit $\mathrm{H}=0$.
 
The uniform spin precession in homogeneous ferromagnetic materials can be described by a torque equation:
\begin{equation}
\frac{d\mathbf{M}}{d t} = -\gamma \mathbf{M}\times\mathbf{H}_{\mathrm{eff}}
\label{torque}\end{equation}
where $\mathbf{H}_{\mathrm{eff}} = -\frac{\partial E}{\partial \mathbf{M}}$, and $E$ is the magnetic free energy of the system, which can be written as $E=-\mathbf{H}\cdot\mathbf{M}+2\pi\mathrm{M}_z^2+K_a{\mathrm{M}_z^2}/{\mathrm{M}_s^2}$. The corresponding phenomenological fields are the demagnetizing field $\mathrm{H}_d=4\pi\mathrm{M}_s$ and the out-of-plane anisotropy field $H_a={2K_a}/{\textrm{M}_s}$. $\textrm{H}_d$ is determined independently by SQUID magnetometry measurements. No in-plane anisotropy field needs to be included since the samples are isotropic within the surface plane, as shown in Fig. 4 for the $n=2$ superlattice. 

\begin{figure}
	\centering
		\includegraphics[width=\columnwidth]{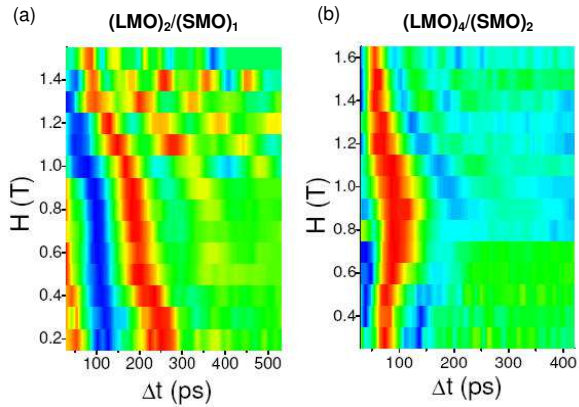}
	\label{fig:fig2}
	\caption{Field dependence of TR-MOKE curves for $n=1$ (a) and $n=2$ (b) superlattices. Red and blue colors correspond to positive and negative precessing M$_z$ components, respectively. An exponential part is subtracted from the raw data to enhance the contrast.}
\end{figure}

The solid lines in Fig. 3 represent fits to the precession frequency using Eq. (1). For the $n=1$ superlattice, both calculated and measured frequencies monotonically decrease with decreasing field to very small values ($< 5$ GHz) at fields below 0.4 T, similar to $\mathrm{La}_{0.67}\mathrm{Sr}_{0.33}\mathrm{MnO}_3$ (LSMO) alloy films~\cite{R12}. An easy plane anisotropy $\mathrm{H}_a = -0.19$ T is obtained for the $n=1$ superlattice, which is comparable to the LSMO alloy film grown on the same $\mathrm{SrTiO}_3$ substrate~\cite{R12}. A small deviation of the measured frequency from the calculated prediction appears at low fields. This discrepancy will be discussed following a description of the $n=2$ superlattice.

For the $n=2$ superlattice, the calculated and measured frequency-field dependence shows very good agreement for applied fields larger than 0.8 T. However, a large discrepancy is revealed at low magnetic fields. The measured frequencies increase with decreasing fields while the calculated curve indicates an opposite trend. For a uniform precession mode, a large frequency at very low fields can only be observed in thin films that are anisotropic in the sample plane. In such films, the in-plane anisotropy field acts like an effective field and provides the torque to the magnetization ~\cite{R13,R16}. However, the ($\mathrm{LMO})_{2n}/(\mathrm{SMO})_n$ superlattices exhibit negligible in-plane anisotropy, as mentioned previously. Nevertheless, a similar effective field must be introduced to account for the large finite frequencies at low fields in the $n=2$ superlattice. 

Since we excluded an anisotropy field, other possible terms contributing to an effective field are bulk dipolar and exchange fields. However dipolar fields can be neglected in our pump-probe TR-MOKE experiments, since dipolar fields only affect spin waves with nonzero wave vector~\cite{R17}. Moreover, exchange torques are not generated for the uniform mode in systems with a single spin population where all coupled spins are parallel and precess with the same amplitude and phase. We thus conclude that a single spin population is not sufficient to describe the field-frequency dependence for the $n=2$ superlattice. In this structure, spins with different characters may emerge due to variations of magnetic interactions within the different layers.

\begin{figure}
	\centering
		\includegraphics[width=\columnwidth]{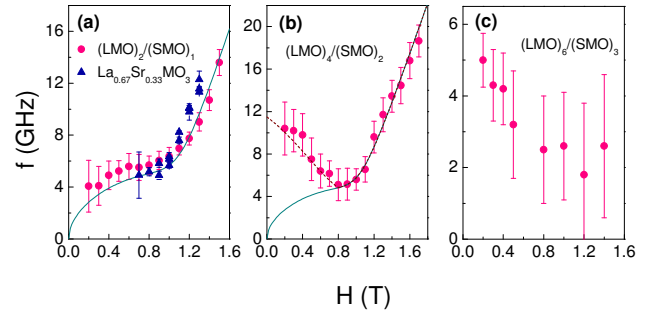}
	\label{fig:fig3}
	\caption{(a)-(c) Field dependence of precession frequency for $n=1$-$3$ superlattices, respectively. The solid lines in (a) and (b) represent calculated frequencies using Eq. (1). The dashed line in (b) shows calculated frequencies with inclusion of an exchange torque.}
\end{figure}

We observe a slight reduction in the magnetic moment per Mn atom in the $n=2$ superlattice, indicating the suppression of ferromagnetic ordering. We suggest that frustrated bonds in regions of varying hole concentration and/or charge carrier mobility result in magnetic disorder and consequently a viscous spin population, as depicted by the blue arrows in Fig. 1(b). In the quasi-static regime, this population is not distinguished from the fast ferromagnetic spins, which constitute a large fraction of Mn sites in the $n=2$ superlattice. A remaining ferromagnetic exchange interaction between viscous and fast spins encourages parallel alignment. Thus, viscous spins rotate with the FM magnetization and applied field; no biased pinning force is produced.  This is evident by the absence of an exchange-biased field, as determined from hysteric magnetization curves shown in Fig. 4. However, we notice an increase in coercivity of the $n=2$ superlattice compared to the alloy film and $n=1$ superlattice, as shown in Fig. 4(a). This may be caused by the resistance of viscous spins during the magnetization reversal process, analogous to the phenomenon in FM/AF heterostructures, where the drag of uncompensated AF spins enhances the coercive field but keeps FM and AF spins aligned along the same direction~\cite{R18}. 

In the dynamic regime, the viscous spins may not readily align with the precessing FM magnetization. FM spins excited by the pump pulse quickly rotate away from the original effective field $\mathbf{H}_\mathrm{eff}$, while viscous spins remain along $\mathbf{H}_{\mathrm{eff}}$ within tens of picoseconds. When the frequency of fast spins (manifest in the magnetization) greatly exceeds the inverse relaxation time of the viscous spins, an exchange torque $\gamma A\mathbf{M}\times\mathbf{M}_v$ is exerted on the precessing magnetization, enhancing the precession frequency. Here, $\mathbf{M}_v$ represents the total magnetization of viscous spins.

 The exchange field $A\mathrm{M}_v$ exerted on to the FM magnetization is calculated to be $\approx0.20$ T to account for the $\approx 11 $GHz precession frequency of the $n=2$ superlattice in the limit $\mathrm{H_{ext}}=0$. For the $n=1$ superlattice under the same condition, we observe a finite but small frequency ($<5$ GHz); similar calculations indicate a much weaker exchange torque ($<0.03$ T) for this system.  
 
 The viscous spins rotate in the same manner as the FM magnetization, but on a much longer time scale ($>10$ ns), hence the exchange fields must be isotropic within the sample plane. This interaction is equivalent to a rotational anisotropy or an "isotropic anisotropy" field~\cite{R19}. A similar anisotropy was observed in spin-glass materials in which spin correlations lead to an additional field parallel to the applied field. For example, a spin-glass like phase can exist at a ferromagnetic/superconductor interface, giving rise to a shift in the resonance field of the FMR spectra relative to that observed in the bulk ferromagnetic material~\cite{R20}. An isotropic FMR shift was also observed in FM/AFM heterostructures~\cite{R21}. Although the origin of these fields remains uncertain, some features are linked to spin-glass like phases.
 
 At zero field, the exchange interaction between viscous spins and fast spins lies within the sample plane. A strong perpendicular field tilts the FM spins out of the sample plane and reduces magnetic disorder as hole hopping is enhanced concurrently. This might generate an ordered FM state within the whole superlattice, as ferromagnetic interaction between Mn spins is promoted. Hence, the exchange torque is diminished at strong applied fields due to a reduction in the number of viscous spins. This process may explain the observed minimum of the precession frequency around 0.8 T for the $n=2$ superlattice. Using this model, we calculated the frequency-field dependence shown by the dashed line in Fig. 3(b). In this calculation, the exchange torque decreases with increasing applied field and vanishes at $\mathrm{H}=0.8$ T. We neglected the field dependence of $\mathrm{H}_d$ and $\mathrm{H}_a$, since the number of viscous spins is much smaller than the number of free spins. This simulation implies a threshold field at which the viscous spins emerge and an inverse linear dependence of viscous spin population on applied field. 
 
The TR-MOKE measurements of the $n=3$ superlattice reveal a similar frequency-field dependence at low fields as for the $n=2$ sample - the precessional frequency increases with decreasing magnetic field, as shown in Fig 3 (c). The same exchange torque generated by viscous spins on the FM magnetization may account for this behavior as well. The overall magnitude of the precession frequency is quite different from that observed in superlattices with $n\leq2$ and LSMO alloy previously discussed. This may be caused by a significant reduction of magnetic moment and demagnetizing field compared to the other structures. The out-of plane anisotropy may also be strongly modified, which dramatically affects the magnitude of the frequency as well. The enhanced damping of the precession in the $n=3$ superlattice is further evidence of the emergence of magnetic disorder. In this superlattice, the large variation of the density of holes may give rise to the coexistence of ferromagnetic and antiferromagnetic phases, which further favors the formation of frustrated bonds at the FM/AF interfaces leading to disordered viscous spins. A strong applied field would align more of these spins, therefore reducing the disorder and exchange torque, similar as in the $n=2$ superlattice.

\begin{figure}
	\centering
		\includegraphics[width=\columnwidth]{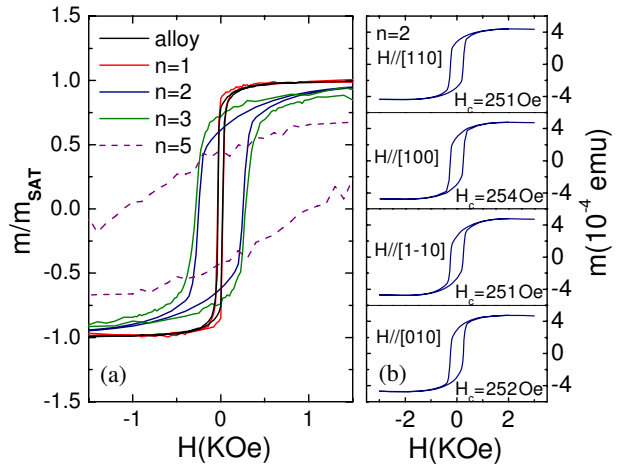}
	\label{fig:fig4}
	\caption{(a) Normalized in-plane M-H loops for superlattices with various periods and an alloy film; (b) M-H loops for an $n=2$ superlattice with field applied along four in-plane principle crystallographic axes.} 
\end{figure}

We do not observe magnetization precession in the $n=5$ superlattice. This may be due to the difficulties of exciting a uniform mode in this structure, where the ferromagnetic order is strongly modulated by a periodic modulation of hole density, as evidenced by neutron scattering experiments~\cite{R8}. Fig. 4(a) shows a very high coercive field ($>1$ kOe) in this superlattice, which may point to pinning of the spins due to coexisting FM and AF regions. The strong frustration at FM and AF interfaces creates a large distribution of pinning fields, giving rise to a broad hysteresis loop. This may preclude a coherent rotation of the FM spins. Furthermore, the modulation of magnetism may cause strong magnon scattering~\cite{R22}, consequently greatly diminishing the dephasing time for the uniform magnetization precession. Thus, we propose that the increased precessional damping is intimately correlated with increasing magnetic disorder, as manifested in large $n$ superlattices.

In conclusion, we have observed coherent spin precession in digital superlattices of $(\mathrm{LMO})_{2n}/(\mathrm{SMO})_n$. The frequency-field dependence for the $n=1$ superlattice is nearly identical to that of random-alloy thin films, consistent with a homogeneous spin character. A negative frequency-field dependence is observed in $n=2$ and $n=3$ superlattices, which results from an exchange torque generated in the dynamic regime between viscous and fast spins. We ascribe the emergence of viscous spins to frustrated bonds and magnetic disorder, which develop upon increasing the superlattice period as indicated by the enhanced damping of the magnetization precession and increased coercive field. Our findings provide new insight into the role of exchange coupling on the fast magnetization dynamics and switching in short-period superlattices composed of two dissimilar magnetic materials.

We gratefully acknowledge financial support by the US Department of Energy, Office of Basic Energy Sciences under contracts DE-FG02-04ER46127 (College of William and Mary), DE-AC02-06CH11357 (Argonne National Laboratory), and DE-AC02-06CH11357 subcontract WO 4J-00181-0004A (University of Illinois).

\end{document}